\documentclass[twoside,twocolumn,english,pra,aps,superscriptaddress]{revtex4-1}
\usepackage[T1]{fontenc}
\usepackage{color}
\usepackage{babel}
\usepackage{textcomp}
\usepackage{amsmath}
\usepackage{amssymb,amsfonts,mathrsfs}
\usepackage{amsthm}
\usepackage{graphicx}
\usepackage[unicode=true,pdfusetitle,
 bookmarks=true,bookmarksnumbered=false,bookmarksopen=false,
 breaklinks=false,pdfborder={0 0 0},backref=false,colorlinks=true]
 {hyperref}
\usepackage{breakurl}

\usepackage{ulem}

\makeatletter
\theoremstyle{plain}

\makeatother

\begin{document}

\preprint{This line only printed with preprint option}

\title{Revealing quantum operator scrambling via measuring Holevo information on digital quantum simulators}

\author{Bin Sun}
\affiliation{Key Laboratory of Atomic and Subatomic Structure and Quantum Control (Ministry of Education),  Guangdong Basic Research Center of Excellence for Structure and Fundamental Interactions of Matter, and  School of Physics, South China Normal University, Guangzhou 510006, China}

\author{Geng-Bin Cao}
\affiliation{Key Laboratory of Atomic and Subatomic Structure and Quantum Control (Ministry of Education),  Guangdong Basic Research Center of Excellence for Structure and Fundamental Interactions of Matter, and  School of Physics, South China Normal University, Guangzhou 510006, China}

\author{Xi-Dan Hu}
\email{huxidan_phys@m.scnu.edu.cn}
\affiliation{Key Laboratory of Atomic and Subatomic Structure and Quantum Control (Ministry of Education),  Guangdong Basic Research Center of Excellence for Structure and Fundamental Interactions of Matter, and  School of Physics, South China Normal University, Guangzhou 510006, China}

\author{Dan-Bo Zhang}
\email{dbzhang@m.scnu.edu.cn}
\affiliation{Key Laboratory of Atomic and Subatomic Structure and Quantum Control (Ministry of Education),  Guangdong Basic Research Center of Excellence for Structure and Fundamental Interactions of Matter, and  School of Physics, South China Normal University, Guangzhou 510006, China}
\affiliation{Guangdong Provincial Key Laboratory of Quantum Engineering and Quantum Materials,  Guangdong-Hong Kong Joint Laboratory of Quantum Matter, and Frontier Research Institute for Physics,\\  South China Normal University, Guangzhou 510006, China}

\date{\today}

\begin{abstract}
Quantum operator scrambling describes the spreading of local operators into the whole system in the picture of Heisenberg evolution, which is often quantified by the operator size growth. Here we propose a measure of quantum operator scrambling via Holevo information of operators, by taking its capacity to distinguish operator information locally. We show that the operator size is closely related to a special kind of Holevo information of operators. Moreover, we propose a feasible protocol for measuring Holevo information of operators on digital quantum simulators based on random states. \textcolor{black}{For the mixed-field Ising model,} our numerical simulations show that the integrable system can be told apart from the chaotic system by measuring the spatial-temporal patterns of Holevo information. Furthermore, we find that error mitigation is required to restore the time-oscillation behavior of Holevo information for the integrable system, a crucial feature distinct from the chaotic one. Our work provides a new perspective to understand the information scrambling and quantum chaos from aspects of Holevo information of operators.

\end{abstract}

\maketitle

\section{Introduction}
In quantum many-body systems, information scrambling provides an important perspective for understanding quantum chaos and has attracted much attention in recent years~\cite{swingle_measuring_2016,patel_quantum_2017,pappalardi_scrambling_2018,qi_measuring_2019,xu_does_2020,ali_chaos_2020,zhao_scaling_2023,alba_eigenstate_2015}. It describes how the local information is scrambled into the whole system via time evolution, distinguishing quantum chaos from integrable systems. Due to its fundamental importance, information scrambling has been widely used to study the dynamics of quantum many-body systems, such as black holes~\cite{hayden_black_2007,guo_distinguishing_2018,bao_holevo_2022,giddings_black_2012}, strange metals~\cite{patel_quantum_2017,banerjee_solvable_2017,ben-zion_strange_2018,hartnoll_theory_2015}, and quantum circuits~\cite{cross_validating_2019,yoshida_disentangling_2019,mi_information_2021}.

There are different measures to quantify information scrambling, including out-of-time-order correlation(OTOCs) and operator size~\cite{rabinovici_krylov_2022,sahu_information_2024,swingle_measuring_2016,zhao_probing_2022,zhang_operator_2023,qi_measuring_2019}. For example, OTOCs have been shown to be sensitive indicators for the detection of quantum chaos, especially in quantum black holes and high-energy physics~\cite{zhao_scaling_2023,huang_chaotic_2020,hayden_black_2007,von_keyserlingk_operator_2018}. Remarkably, recent rapid advances in quantum processors enable us to access OTOCs experimentally, which has allowed the research to broaden to the experimentally verifiable stage. On the other hand, information scrambling can also be measured via operator size spreading~\cite{qi_measuring_2019,khemani_operator_2018,nahum_operator_2018,weisse_operator_2024,parker_universal_2019,zhao_scaling_2023,zhao_super-exponential_2021,zhao_superexponential_2020,hu_quantum_2023,zhang_dynamical_2023,zhang_operator_2023,gao_scrambling_2024}. In this perspective, a simple local operator after time evolution will spread into the whole system and become a linear combination of nonlocal operators. According to this description, operator size growth can intuitively characterize quantum chaotic systems with a saturation behavior that distinguishes them from integrable systems. Moreover, the operator size can be obtained in quantum simulators by Bell measurements~\cite{hu_quantum_2023,weisse_operator_2024}. However, previous studies have lacked a quantum information perspective related to the degree of operator distinguishability for describing the occurrence of information scrambling.

Another potentially possible measure to characterize quantum information scrambling can be based on Holevo information. Holevo information originates as a measure of the upper bound on the accessible information between two players when Bob tries to receive information from Alice~\cite{holevo_bounds_1973,qi_holevo_2022,nielsen_quantum_2010,yuan_quantum_2022,hayden_black_2007,guo_distinguishing_2018,holevo_coding_1998}. As a fundamental quantum information theoretic quantity, Holevo information gives rise to a new perspective to understand quantum many-body dynamics, ranging from visualization of quantum many-body scars~\cite{yuan_quantum_2022}, investigation of phase transitions in black holes~\cite{bao_holevo_2022}, and characterizing the memory of many-body localization~\cite{Nico-2022}. While those studies exploit the Holevo information of quantum states, it wonders whether the Holevo information can be generalized to operators and thus can be utilized to probe the scrambling of operators.

In this paper, we propose a measure of quantum operator scrambling with Holevo information dynamics and reveal its connection to the operator size. We propose a feasible protocol for measuring Holevo information. The scheme is based on a mapping between Pauli operators and Bell states and utilizes the randomized channel-state duality, which saves nearly half the number of qubits. We demonstrate the effectiveness of the algorithm by using a mixed-field Ising model. By numerical simulations, we show that spatial-temporal patterns of Holevo information are quite different between the integrable system and the chaotic system. However, the behaviors become close once there are quantum noises, which are inevitable on real quantum hardware. Nevertheless, by error mitigation, we find the time-oscillation behavior of Holevo information for the integrable system, a crucial feature distinct from the chaotic one, can be clearly restored. Our work provides a new perspective to understand and probe the information scrambling and quantum chaos via Holevo information dynamics of operators.

The paper is organized as follows. In Section ~\ref{s2}, we review the definition of operator size and give the definition of Holevo information of operator, and then we discuss some connections between them. In Section ~\ref{s3}, we improve the original algorithm based on state-channel duality by using random states. In Section ~\ref{s4}, we give simulation results with corresponding analysis and show the effectiveness of error mitigation
in the presence of quantum noises. Finally, we give conclusions in Section ~\ref{s5}.

\section{operator size and holevo information of operator }
\label{s2}

In this section, we first review the definition as well as the meaning of operator size and Holevo information of operators for quantum systems. Then, we establish some connections between them.

\subsection{Operator Size}

In the Heisenberg picture, simple operators can be evolved into linear superpositions of increasingly complicated operators~\cite{huang_chaotic_2020,klug_chaos_2019,roberts_chaos_2017,huang_lieb-robinson_2018}. For lattice systems, those complicated operators are non-local in the sense of having support over many different sites. One can endue for an averaged size for an operator, which characterizes the scrambling of a local operator into a superposition of many nonlocal ones.

The evolution of an operator is governed by the Heisenberg equation,
\begin{equation}\label{Heisenberg}
\frac{d\hat{O}(t)}{dt}=i[\hat{H},\hat{O}(t)],
\end{equation}
whose solution is: $\hat{O}(t)=e^{i\hat{H}t}\hat{O}(0)e^{-i\hat{H}t}$.

By properly selecting an orthonormal basis of Hermitian operators $\{\hat{O}_\alpha\}$ with the orthonormal condition $2^{-N}\operatorname{tr}(\hat{O}_{\alpha}\hat{O}_{\beta})=\delta_{\alpha\beta}$, such as the Pauli basis, every operator in a generic $N$ qubit lattice system can be decomposed on this basis. Specifically, each Pauli basis can be expressed as a product of Pauli operators. Here, the Pauli matrices are defined as $\hat{\sigma}^0=\hat{I}$, $\hat{\sigma}^1=\hat{X}$, $\hat{\sigma}^2=\hat{Y}$, and $\hat{\sigma}^3=\hat{Z}$. The operator $\hat{O}(t)$ can then be expanded as,
\begin{equation}\label{eq:expand_O}
\hat{O}(t)=\sum_{k=0}^{4^{N}-1}C_{k}(t)\hat{P}_{k}, \quad \hat{P}_{k}=\bigotimes_{n=1}^N\hat{\sigma}^{k_n}_n,
\end{equation}
where $k=k_1k_2\dots,k_N$ with $k_n \in \{0,1,2,3\}$, and $\hat{\sigma}^{k_n}_n$ denotes the Pauli operator acting on site $n$. The coefficient can be evaluated as,
\begin{equation}
C_{k}(t)=2^{-N}\text{tr}[\hat{O}(t)\hat{P_{k}}].
\end{equation}

Notice that $\hat{O}(t)$ in Eq.~\eqref{eq:expand_O} satisfies the following normalization conditions $2^{-N}\text{tr}(\hat{O}(t)^2)=1$. Due to orthogonality under Pauli basis vectors, $2^{-N}\text{tr}(\hat{P}_k\hat{P}_j)=\delta_{kj}$, it is not difficult to deduce that the coefficient sum of squares is 1,
\begin{equation}\label{Ckt}
\begin{split}
2^{-N}\text{tr}(\hat{O}(t)^2) = & \sum_{k=0}^{4^N-1} \sum_{j=0}^{4^N-1} C_{k}(t) C_{j}(t) 2^{-N} \text{tr}(\hat{P}_k \hat{P}_j)\\
= & \sum_{k=0}^{4^N-1} \left| C_{k}(t) \right|^2 =1.
\end{split}
\end{equation}

In other words, the squared moduli of the expansion coefficients of an operator under the Pauli operator basis can be interpreted as probabilities. The averaged operator size for $\hat{O}(t)$ can be expressed as,
\begin{equation}\label{OS} 
L(\hat{O}(t))= \sum_{k=0}^{4^N-1}\left|C_{k}(t)\right|^2\times l(\hat{P}_k).
\end{equation}
Here $l(\hat{P}_k)$ is the operator size of Pauli basis $\hat{P}_{k}$, which counts the number of non-$\hat{I}$ operators in the Pauli string~(or the number of non-zeros in the quaternary number $k=k_1k_2...k_N$, e.g., $l(\hat{\sigma}^0)=0$, $l(\hat{\sigma}^1)=l(\hat{\sigma}^2)=l(\hat{\sigma}^3)=1$). The averaged operator size can also be written as a summation of local operator densities $L_{n}$ at the lattice site $n$~\cite{hu_quantum_2023}, \begin{equation}\label{L_sum}
L(\hat{O}(t))=\sum_{n=1}^{N}L_{n}, \quad L_{n}=\sum_{k_n=0}^{3}|C_{k_n}(t)|^{2} \, l(\hat{\sigma}^{k_n}),
\end{equation}
where $C_{k'_n}(t)=\sum_{\text{if }k_n=k'_n}[2^{-N}\text{tr}(\hat{O}(t)\hat{P}_{k})]$ with $k=k_1k_2...k_n..k_N$. One can verify that $\sum_{k_n=0}^{3}|C_{k_n}(t)|^{2}=1$, thus explaining $|C_{k_n}(t)|^{2}$ as the probability of finding operator $\hat{\sigma}^{k_n}$ at the site $n$.

Eq.~\eqref{OS} and Eq.~\eqref{L_sum} allow us to quantify the size of an operator in a many-body quantum system based on its decomposition in the Pauli basis and provide insight into the system's dynamical behavior, particularly in the context of quantum chaos and operator spreading.

\subsection{Holevo Information of Operators}
In quantum information theory, Holevo information, or Holevo bound $\chi$, is a measure of the maximum accessible information between two separate agents~\cite{holevo_bounds_1973,nielsen_quantum_2010}. Let Alice send $\rho_{X} $ to Bob, where $\rho_{X}$ is a set of density matrices $\{\rho_{1},\rho_{2},\cdots,\rho_{R}\}$ with probabilities $\{p_{1},p_{2},\cdots,p_{R}\}$ respectively.
By positive operator-valued metrics (POVMs) with outcome $Y$, Bob can access information upper bound by $\chi$,
\begin{equation} \label{eq:chi}
I(X,Y)\leq\chi(\rho_{X})=S\left(\sum_{i=1}^{R}p_{i}\rho_{i}\right)-\sum_{i=1}^{R}p_{i}S\left(\rho_{i}\right),
\end{equation}
where $I(X,Y)$ is the mutual information between $X$ and $Y$, and $S(\rho)=-\text{tr}(\rho\ln{\rho})$ is the von Neumann entropy. Since Holevo information can characterize the distinguishability of states, it is naturally used to characterize quantum information scrambling dynamics~\cite{yuan_quantum_2022,guo_distinguishing_2018}.

Here, we would like to extend Holevo information for operators and further explore it to characterize quantum operator scrambling. Our aim is to define the Holevo information of operators $\hat{O}_X$ in a set $\{\hat{O}_1,\hat{O}_2,\dots,\hat{O}_R\}$ with probabilities $\{p_{1},p_{2},\cdots,p_{R}\}$ respectively. For this, we exploit a channel-state duality~\cite{jiang_channel-state_2013,yan_randomized_2022,choi_completely_1975}, which can directly map an operator to a quantum state. Then the original definition of Holevo information may be directly applied. For a unitary and Hermitian operator $\hat{O}$ considered in this paper, the channel-state duality can be written as~\cite{leifer_quantum_2006,jiang_channel-state_2013,lancien_approximating_2024},
\begin{equation}\label{eq:channel-state-duality}
    \hat{O} \longrightarrow |\hat{O}\rangle \equiv\hat{O}\otimes \hat{I} \prod_i|B^0_i\rangle.
\end{equation}
Here $|B^0_n\rangle=\frac{1}{\sqrt{2}}\left(|0_n0_{n'}\rangle+|1_n1_{n'}\rangle\right)$ is the Bell state, $\hat{O}$ performs on the system qubit (indexed as $n=1\dots N$ and $n'$ indexes the ancillary qubit). For later use, we define three other Bell basis,
\begin{equation}\label{BellStPpt}
\begin{split}
|B^1_n\rangle=\hat{X}_n|B^0_n\rangle=\frac{1}{\sqrt{2}}\left(|1_n0_{n'}\rangle+|0_n1_{n'}\rangle\right),\\
|B^2_n\rangle=\hat{Y}_n|B^0_n\rangle=\frac{i}{\sqrt{2}}\left(|1_n0_{n'}\rangle-|0_n1_{n'}\rangle\right),\\
|B^3_n\rangle=\hat{Z}_n|B^0_n\rangle=\frac{1}{\sqrt{2}}\left(|0_n0_{n'}\rangle-|1_n1_{n'}\rangle\right).\\
\end{split}
\end{equation}
The operator $\hat{O}$, originally as a linear combination of Pauli strings, now becomes a state as a superposition of Bell basis $\hat{P}_k\otimes \hat{I} \prod_i|B^0_i\rangle$.

By such channel-state duality, the Holevo information of operators is turned into the Holevo information of states, namely a set of states $\{|\hat{O}_{1}\rangle,|\hat{O}_{2}\rangle,\cdots,|\hat{O}_{R}\rangle\}$ with probabilities $\{p_{1},p_{2},\cdots,p_{R}\}$ respectively. The calculation of Holevo information of operators thus can follow Eq.~\eqref{eq:chi}.

We further define Holevo information to distinguish operators locally. This is necessary as two initially orthogonal Pauli operators will always be orthogonal to each other with the same Hamiltonian evolution, e.g., they can always be distinguished globally. Thus, to reveal the operator scrambling, one has to measure the difference of the two operators only locally. With the channel-state duality, one can use the reduced density matrices on the region $A$ by tracing out the complementary region $\bar{A}$~(both system qubits and ancillary qubits in regime $\bar{A}$),
\begin{equation} \rho^A_{\hat{O}}=\text{tr}_{\bar{A}}|\hat{O}\rangle\langle\hat{O}|.
\end{equation}
By tracing out some degrees of freedom, the two states $\rho^A_{\hat{O}_1}$ and $\rho^A_{\hat{O}_2}$ become less distinguishable than the two states $|\hat{O}_1\rangle$ and $|\hat{O}_2\rangle$.

To be concrete for studying operator scrambling, we consider two different Pauli operators located at site $m$, namely $\hat{O}_1(0), \hat{O}_2(0) \in \{\hat{I}_m,\hat{X}_m,\hat{Y}_m,\hat{Z}_m\}$. After Heisenberg evolution of time $t$, the two time-evolved operators are $\hat{O}_1(t)$ and $\hat{O}_2(t)$. We reveal the scrambling of information at site $n$ by Helevo information of local operators, which turns to be Helevo information of a set of states $\{\rho^n_{\hat{O}_1(t)},\rho^n_{\hat{O}_2(t)}\}$ with probabilities $\{\frac{1}{2},\frac{1}{2}\}$. We denote the Helevo information as,
\begin{eqnarray}\label{Holevoinformation_n}
    &&\chi_n(\hat{O}_1(t), \hat{O}_2(t)) =\nonumber \\
    &&S\left( \frac{\rho^n_{\hat{O}_1(t)} + \rho^n_{\hat{O}_2(t)}}{2} \right) - \frac{S\left(\rho^n_{\hat{O}_1(t)}\right) + S\left(\rho^n_{\hat{O}_2(t)}\right)}{2}.
\end{eqnarray}

An interesting relation between the Holevo information and the operator densities can be revealed as follows.
Let the probabilities of finding the four Pauli operators at the site $n$ for the operator $\hat{O}(t)$ as \(P(\hat{I}) = P_0\), \(P(\hat{X}) = P_1\), \(P(\hat{Y}) = P_2\), and \(P(\hat{Z}) = P_3\). Then, the operator density at site $n$ is given by:
\begin{equation}\label{L_i}
L_{n} = \sum_{i=1}^{3} P_{i} = 1 - P_{0}.
\end{equation}

On the other hand, we can consider the Holevo information of $\chi_n(\hat{I},\hat{O}(t))$, which describes how a time-evolved operator $\hat{O}(t)$ differs from the background~(identity operators) at each site. Note that the reduced density matrix after channel-state duality at the site $n$ for $\hat{I}$ is $B^0_n=|B^0_n\rangle\langle B^0_n|$. The Holevo information can be expressed by the four probabilities, $\{P_i\}$, which turns out to be,
\begin{equation}
\begin{split}
\chi_n(\hat{I},\hat{O}(t))=&S\left(\frac{B^0_n+\rho^n_{\hat{O}(t)}}{2}\right)-\frac{S\left(\rho^n_{\hat{O}(t)}\right)}{2}-\frac{S(B^0_n)}{2}\\
=&-\frac{P_{0}+1}{2}\log_{2}{\frac{P_{0}+1}{2}}-\sum_{k_n=1}^{3}\frac{P_{k_n}}{2}\log_{2}{\frac{P_{k_n}}{2}}\\
&+\sum_{k_n=0}^{3}\frac{P_{k_n}}{2}\log_{2}{P_{k_n}}\\
=&\frac{1}{2}\left[2+\log_{2}{\frac{P_{0}^{P_{0}}}{(P_{0}+1)^{P_{0}+1}}}\right].
\end{split}
\end{equation}

Thus, we can establish a functional relationship between \(\chi_n\) and \(L_{n}\) according to Eqs.~\eqref{L_i} and the above equation,
\begin{equation}\label{chivs}
\chi_n(L_{n}) = \frac{1}{2} \left( 2 + \log_{2} \left[ \left( \frac{1 - L_{n}}{2 - L_{n}} \right)^{1 - L_{n}}  \frac{1}{2 - L_{n}} \right] \right).
\end{equation}
The Holevo information of local operators is a monotonicity function of the operator density, as the derivative of $\chi_n$ to the operator density $L_n$ is always positive for $L_n\in [0,1]$. In this regard, operator density can be viewed as a special kind of Holevo information for local operators.

\section{Protocol for measuring Holevo information of operators}
\label{s3}
In this section, we give a protocol for measuring Holevo information of local operators on digital quantum simulators by using random states.

As already investigated in Ref.~\cite{hu_quantum_2023}, the operator size of a system of $N$ qubits can be evaluated with Bell measurements, which should use $2N$ qubits by the channel-state duality. Such a protocol can be directly implemented for measuring the Holevo information. However, as only two qubits~(a system qubit and a corresponding ancillary qubit) are required to be measured each time, the other ancillary qubits are traced out initially. In this regard, the initial state can be prepared with one Bell state and a maximal mixed states of $N-1$ qubits. Moreover, the maximal mixed states can be approximated and realized with a few random states~\cite{cross_validating_2019,harrow_approximate_2023}.  By replacing Bell states with random states, the protocol can exploit the randomized channel-state duality for measuring operator size and Holevo information with Bell measurements, which reduces the number of qubits from $2N$ to $N+1$.

\begin{figure}[htbp] \centering
	\includegraphics[width=8.8cm]{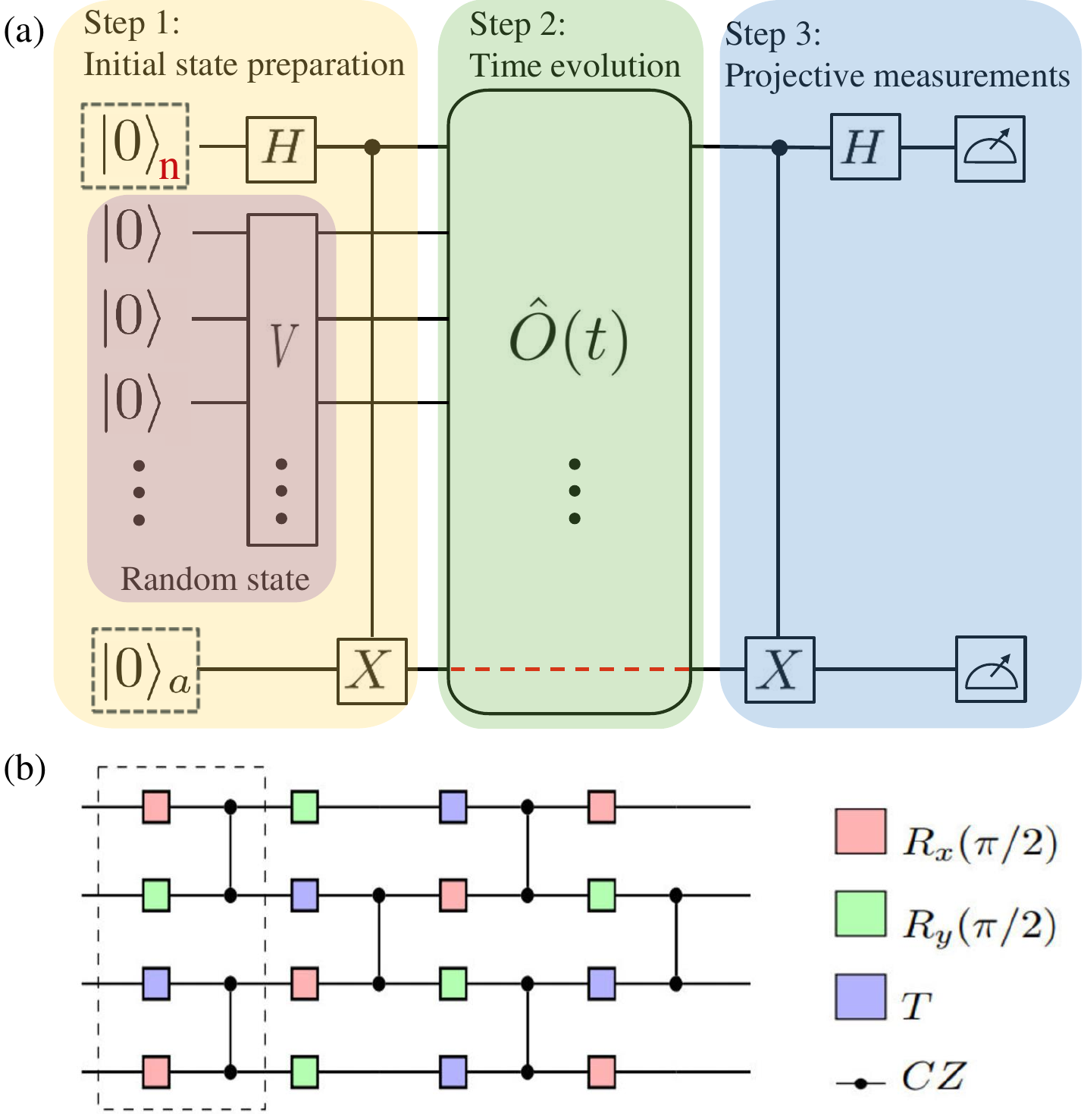}
	\caption{(a) The quantum circuit for measuring the Holevo information of operators at the site $n$ consists the three steps: the initial state preparation with $V$ for generating the random states, the time evolution, and the projective measurements. \textcolor{black}{This circuit should be executed $M$ times with random $V$ to approximate the maximal mixed state.} (b) The sub-circuit, \textcolor{black}{an instance of the $V$}, for the random unitary $V$ consists of $d$ blocks, each block having alternating layers of single-qubit and two-qubit quantum gates. \textcolor{black}{The Controlled-Z (CZ) gate effectively introduces a phase difference to establish entanglement between adjacent qubits.}}\label{circuit}
\end{figure}

We briefly give the implementation of the protocol. We choose a  sub-circuit for random state preparation, which refers to a one-dimensional lattice~\cite{harrow_approximate_2023}. The protocol for measuring the Holevo information of operators at site $n$ consists of four main steps~(also illustrated in Fig.~\ref{circuit}).

\emph{Step 1: Initial state preparation.}
Prepare the initial state being the Bell state at the site $n$, and the rest is set to the maximal mixed state,
\begin{equation}\label{ISC} 
	\rho_0 = |B^0_n\rangle\langle B^0_n| \otimes \frac{\hat{I}}{2^{N-1}}.
\end{equation}
The maximal mixed state $\frac{\hat{I}}{2^{N-1}}$ can be approximated with an ensemble average of $M$ random states generated by \textcolor{black}{executing the random circuit $M$ times} in Fig.~\ref{circuit} (b), which can be written as,
\begin{equation}\label{EOP} 
 \frac{\hat{I}}{2^{N-1}}\approx \frac{1}{M}\sum_{k=1}^{M}\left|\psi_{k}\right\rangle\left\langle\psi_{k}\right|.
\end{equation}
We give an analysis of the required $M$ in Sec.~\ref{s4} for accurate approximating of the maximal mixed state, which shows that $M$ is almost independent of the system size $N$ under the same accuracy~\cite{yan_randomized_2022,vairogs_extracting_2024}. \textcolor{black}{Here, the random unitary $V$ consists of $d$ blocks, each block consisting of two layers. For the first layer, each qubit is randomly rotated on its own X, Y, or Z axis, which can be realized as a single-qubit gate as shown in Fig.~\ref{circuit}(b). Then, the second layer connects qubits by two-qubit Controlled-Z quantum gates.}

\emph{Step 2: Time evolution.} According to Eq.~\eqref{Heisenberg} and its solution, we should perform the unitary operator $\hat{O}(t)=e^{i\hat{H}t}\hat{O}(0)e^{-i\hat{H}t}$ on the initial state $\rho_0$, and one gets,
\begin{equation}\label{FSC} 
        \begin{split}
		\rho(t)=&\hat{O}(t)\rho_0 \hat{O}^\dagger(t).\\
	\end{split}
\end{equation}
In digital quantum simulation, the Hamiltonian evolution operator $e^{\pm i\hat{H}t}$ should be decomposed into single-qubit and two-qubit quantum gates with Trotter decomposition~\cite{suzuki_general_1991,trotter_product_1959}.

\emph{Step 3: Projective measurements.} Perform the Bell measurements to get the probabilities, $P_{i}=\text{tr}(\rho(t)|B^i_n\rangle\langle B^i_n|)$ for $i=0,1,2,3$. The projective measurements should be repeated until significant accurate probabilities are obtained.

\emph{Step 4: Calculate the Holevo information.} After obtaining the probabilities $\{\{P_{1i}\},\{P_{2i}\}\}$ corresponding to $\{\hat{O}_{1}(t), \hat{O}_{2}(t)\}$ respectively, according to the Eq.~\eqref{Holevoinformation_n} , we can get the expression of Holevo information,
\begin{equation}\label{FSC} 
        \begin{split}
\chi_n(\hat{O}_{1}(t), \hat{O}_{2}(t)) =&\sum_{i=0}^{3}\frac{(P_{1i}+P_{2i})}{2}\log_{2}{\frac{(P_{1i}+P_{2i})}{2}}\\&-\sum_{i=0}^{3}\frac{(P_{1i}\log_{2}{P_{1i}}+P_{2i}\log_{2}{P_{2i}})}{2}.
    \end{split}
\end{equation}

\section{Simulation results}
\label{s4}
In this section, we show the simulation result of the Holevo information of operators using a model Hamiltonian, the mixed-field Ising model. We first give an analysis of the number of random states required to accurately approximate the mixed state. The spatial-temporal patterns of Holevo information for the model with both integrable and chaotic behaviors are given. Finally, we investigate the effects of quantum noise as well as incorporate error mitigation for measuring the Holevo information.

The mixed-field Ising model is a prototypical model for investigating distinct dynamics and has been studied in various contents~\cite{kuelske_ising_2025,nakamura_surface_2024,suzuki_quantum_2013}. Its Hamiltonian can be written as:
\begin{equation}\label{MFIM}
\hat{H}=J\sum_{n=1}^{N-1}\hat{Z}_{n}\hat{Z}_{n+1}+h_x\sum_{n=1}^N\hat{X}_n+h_z\sum_{n=1}^N\hat{Z}_n.
\end{equation}

The model can exhibit both integrable and chaotic regimes depending on the presence or absence of longitudinal field ($h_z$ term)~\cite{von_keyserlingk_operator_2018,chan_spectral_2021,hu_quantum_2023}, respectively. Here, we chose $J=1$ and $h_x=1$. When $h_z=0$, the mixed-field Ising model becomes the well-known transverse field Ising model, which has $Z_2$ symmetry. In contrast, when $h_z\neq0$, the model has no $Z_2$ symmetry.

\begin{figure}[htbp] \centering
        \hspace*{0cm}
	\includegraphics[width=8.8cm]{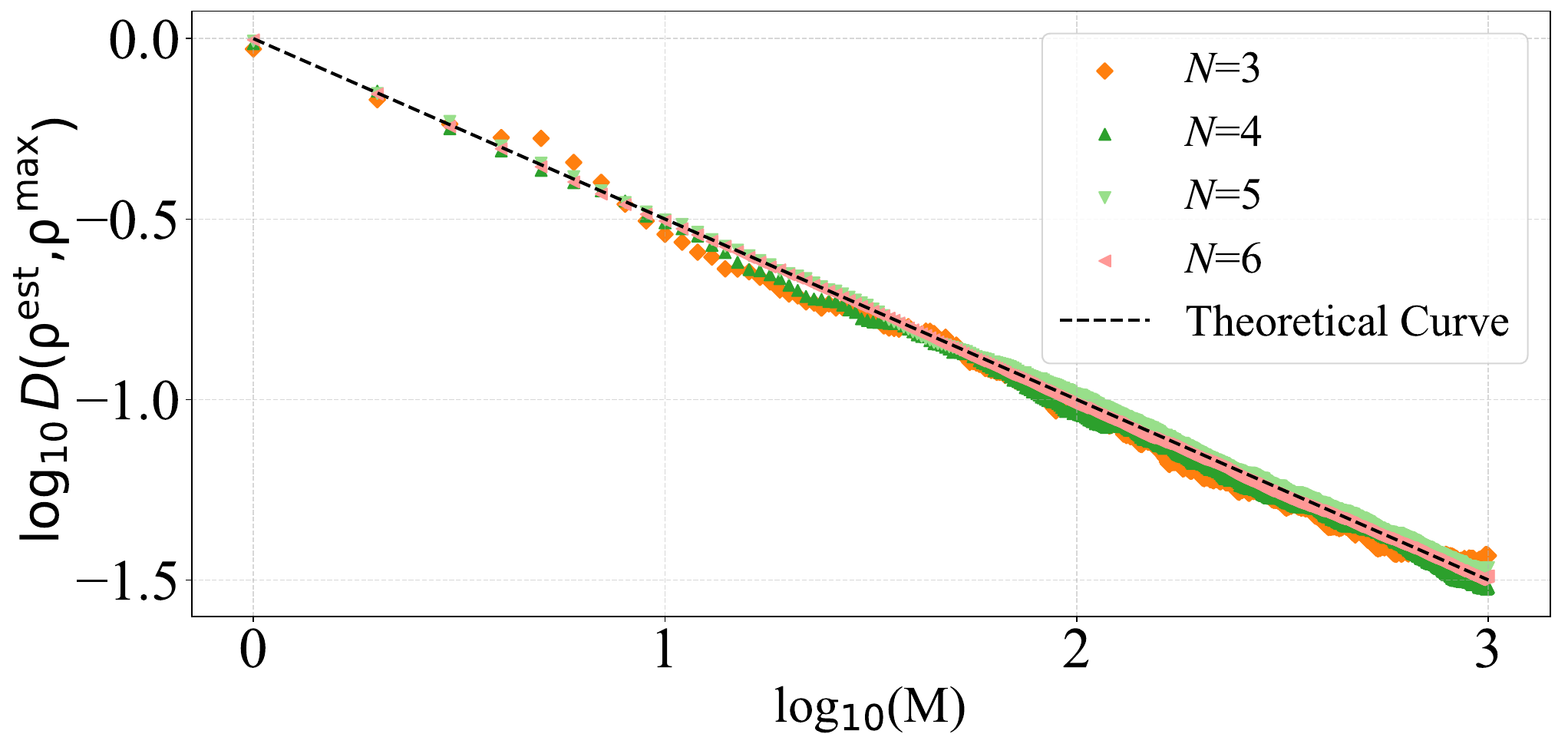}
	\caption{The trace distance $D(\rho^{est},\rho^{max})$ measures the accuracy of using an ensemble average of random states to approximate the maximal mixed state. Here, $N$ is the system size, and $M$ is the number of random states.}\label{trace distacne}
\end{figure}

For measuring the properties of operators, we refer to a randomized channel, where the maximal mixed state is approximated with an ensemble of random states~\cite{richter_simulating_2021,boixo_characterizing_2018}. To make the protocol efficient, the number of random states should not be large. By numeral investigation, we show that the required number of random states for accurately approximating the maximal mixed state is essentially independent of the system sizes, which agrees with Ref.~\cite{yan_randomized_2022}. Typically, in a one-dimensional lattice, the number of blocks required for preparing a $1$-design random state~(its ensemble average is the maximal mixed state) scales linearly with the system size $N$,  $d\sim O(N)$~\cite{harrow_approximate_2023}. In our simulation with a system size $N=7$, we chose $d=8$ as the number of blocks to prepare the random state.

By taking an ensemble average of $M$ random states, we have,
\begin{equation}\label{EOP} 
\rho^{est}=\frac{1}{M}\sum_{k=1}^{M}\left|\psi_{k}\right\rangle\left\langle\psi_{k}\right|.
\end{equation}
The accuracy of the approximation can be characterized by the trace distance between $\rho^{est}$ and the maximum mixed state $\rho^{max}=\hat{I}/2^{N-1}$ of $N-1$ qubits,
\begin{equation}\label{EFSR}
D(\rho^{est}, \rho^{max}) = \frac{1}{2} |\rho^{est} - \rho^{max}|_1.
\end{equation}
As shown in Fig.~\ref{trace distacne}, we find that the trace distance is independent of the system size $N$ and relies on the number of random states as $\frac{1}{\sqrt{M}}$.


\begin{figure}[htbp] \centering
\includegraphics[width=8.9cm]{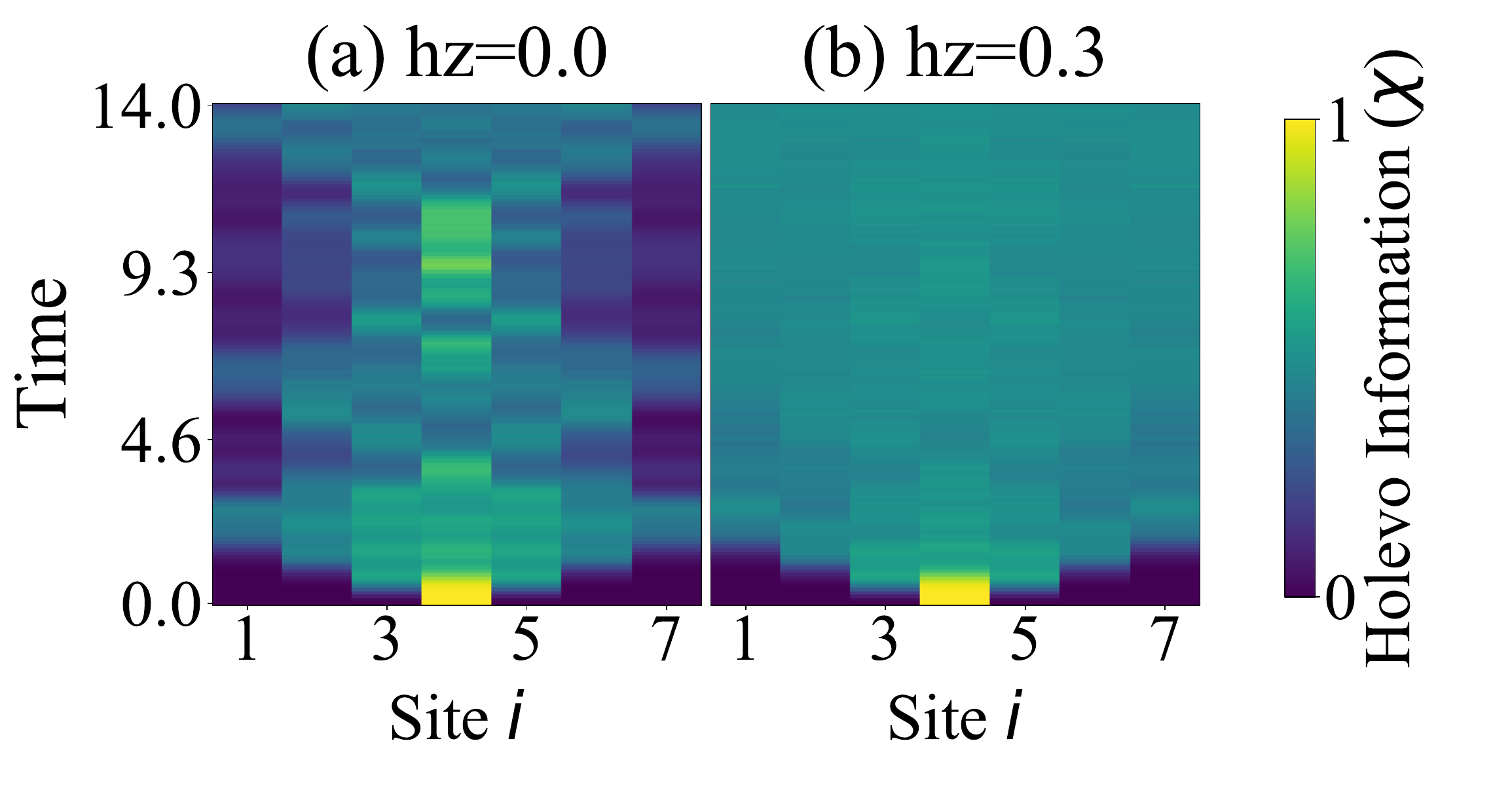}
\caption{(Color online). Spatial-temporal patterns of Holevo information $\chi_n(\hat{I},\hat{X}_4(t))$ for (a) $h_z=0$ and (b) $h_z=0.3$. The system size is $N=7$, with an initial operator $X$ set in the middle, $j=4$. }\label{Holevo informationIX}
\end{figure}

\begin{figure}[htbp] \centering
\includegraphics[width=8.9cm]{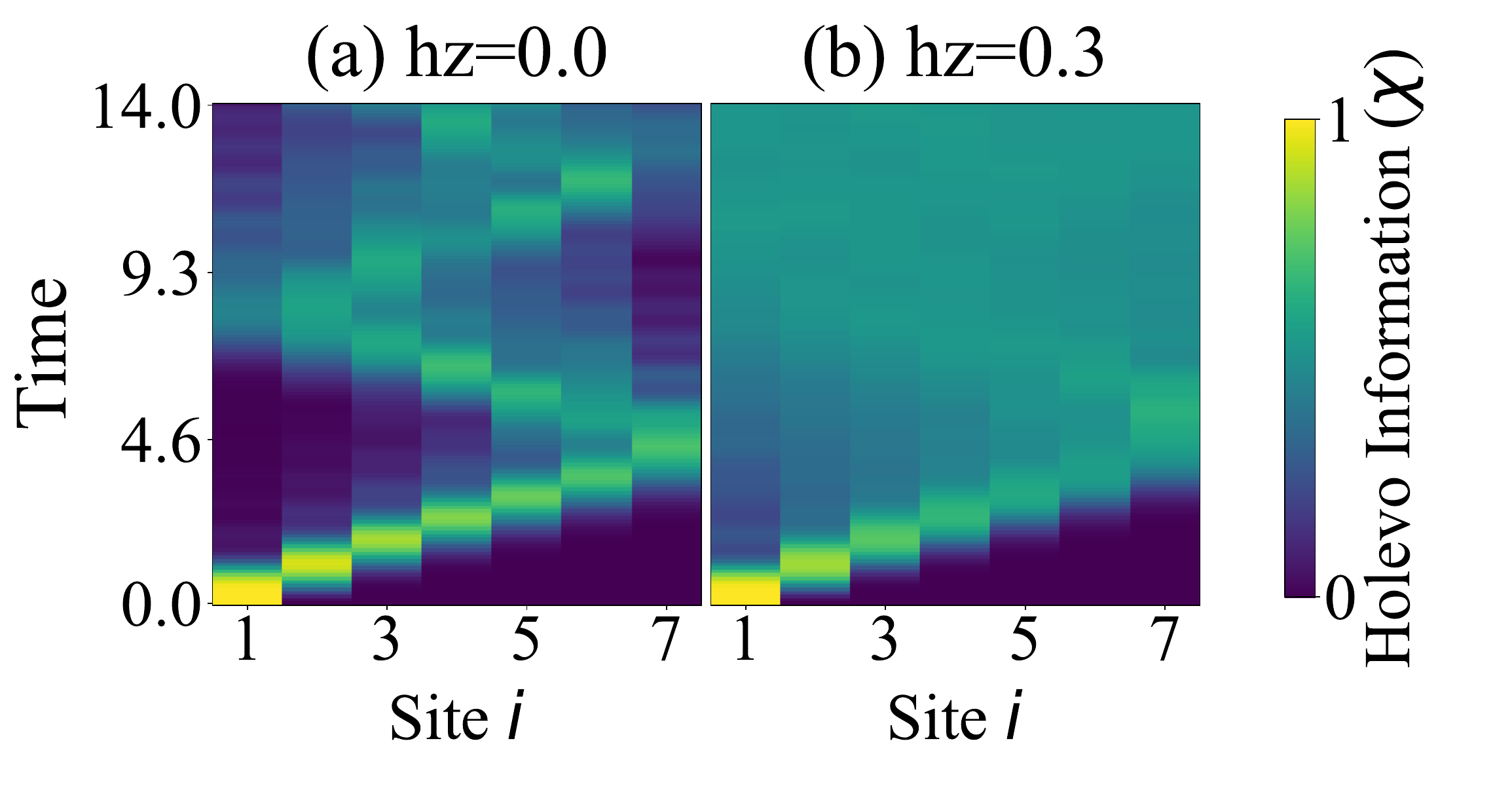}
\caption{(Color online). \textcolor{black}{Spatial-temporal patterns of Holevo information $\chi_n(\hat{I},\hat{X}_1(t))$ for (a) $h_z=0$ and (b) $h_z=0.3$. The system size is $N=7$ with an initial operator $X$ set at the leftmost, $j=1$. }}\label{Holevo informationIX_site1}
\end{figure}

\begin{figure}[htbp] \centering
\includegraphics[width=8.9cm]{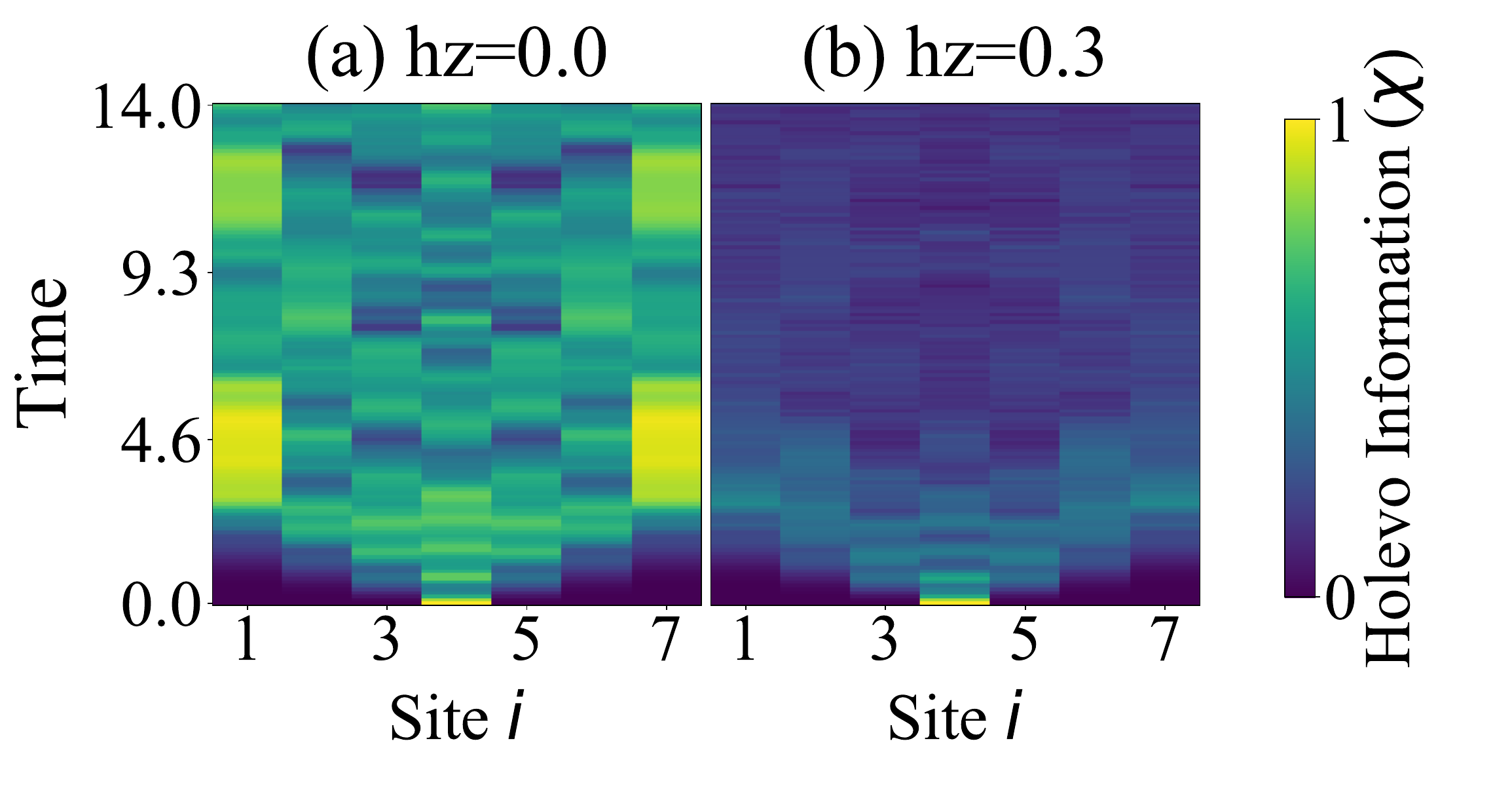}
\caption{(Color online). Spatial-temporal patterns of Holevo information $\chi_n(\hat{X}_4(t),\hat{Y}_4(t))$ for (a) $h_z=0$ and (b) $h_z=0.3$. The system size is $N=7$.} \label{Holevo informationXY}
\end{figure}

\begin{figure*}[t]
        \centering    \includegraphics[width=0.9\linewidth]{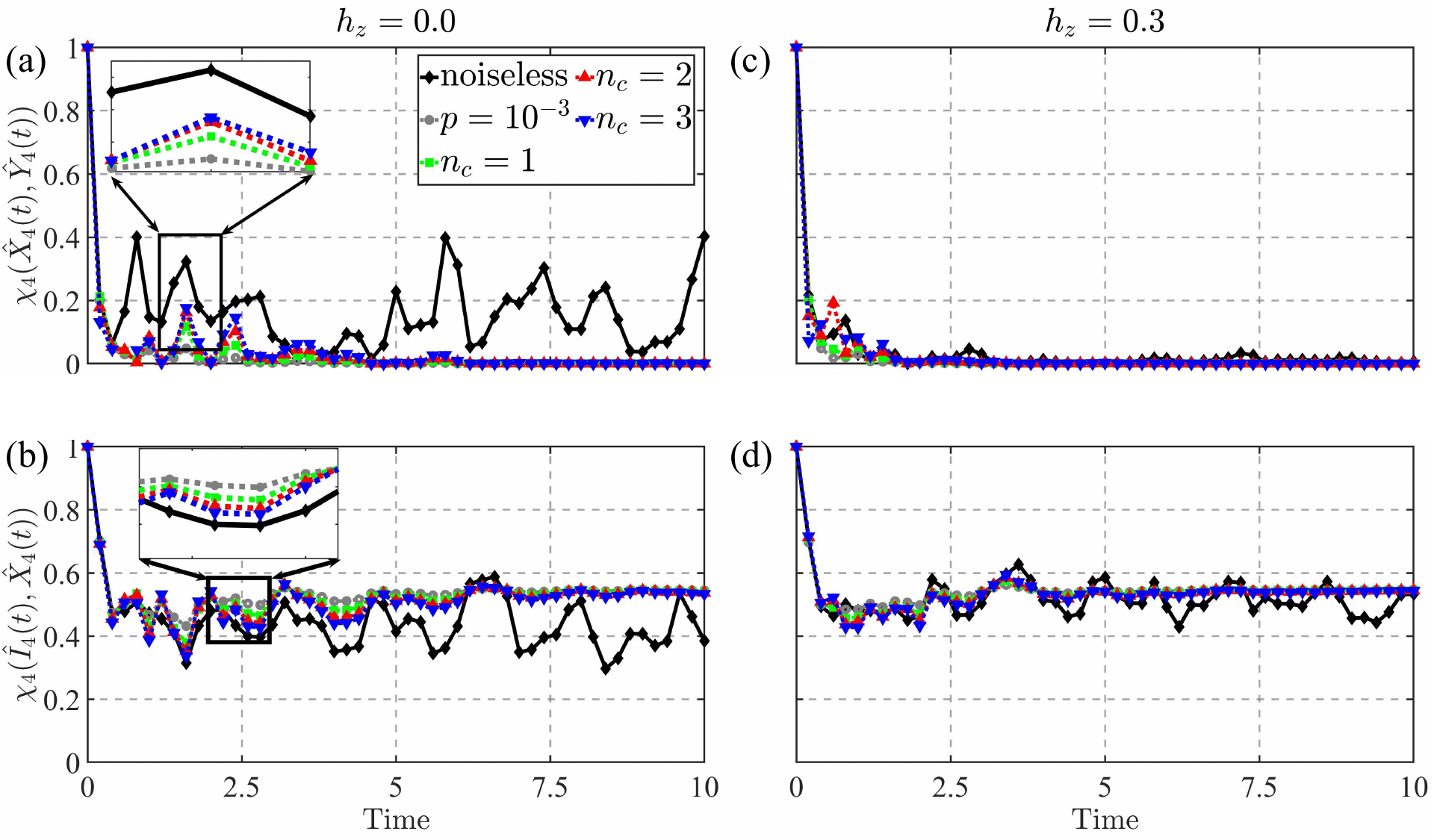}
    \caption{Effects of quantum noises and error mitigation for evolution of Holevo information of operators with time. The noise rate is $p=10^{-3}$, and the error mitigation by Richardson's deferred method is taken for orders up to $n_c=3$. (a), (b) show results for the integrable system, with subplots of small regions for easy viewing. (c), (d) show results for the chaotic system. The system size is $N=7$. \textcolor{black}{The initial operator $X$ is set at the center of the chain, $j=4$}.}
    \label{error}
\end{figure*}

We precede to investigate the quantum operator scrambling from the spatial-temporal pattern of Holevo information. The system size is $N=7$. The Heisenberg evolution of an initial local Pauli operator is performed by Trotter decomposition of the Hamiltonian evolution $e^{\pm i \hat{H}t}$, with a step size $0.1$.  We first give Holevo information $\chi_n(\hat{I}_j(t),\hat{X}_j(t))$, which characterizes the scrambling of Pauli operator $\hat{X}$ at the site $j$ into another site $n$ after a time evolution of $t$. Here $\hat{I}_j(t)=\hat{I}$ is taken as a background. The initial operator is set in the middle, $j=4$. The results are presented in Fig.~\ref{Holevo informationIX}. It can be clearly seen that spatial-temporal patterns are distinct for the integrable ($h_z=0$) and chaotic situations ($h_z=0.3$). \textcolor{black}{Along the light-cone structure, information propagates lastingly in the integrable case and exhibits oscillatory behavior by bouncing from the boundary, while it is rapidly decaying in the chaotic one.} Moreover, in the chaotic case, Holevo information of all sites will become uniform but not zero, which corresponds to the scrambling of local operators into the whole system. Since Holevo information $\chi_n(\hat{I}_j(t),\hat{X}_j(t))$ is closely related to the operator density, we may also explain the results from the aspect of operator density spreading. For the chaotic case, an initial operator at the center spreads to the whole system until the density of operators at each site $L_{n}=
\frac{D^{2}-1}{D^2}$~\cite{qi_measuring_2019}, where $D$ is the dimension of the Hilbert space at each site. For our case $D=2$, $L_n=0.75$. According to Eqs.~\eqref{chivs}, the corresponding Hovelo information tend to be $\chi_{n}(\hat{I}(t),\hat{X}(t))= 0.54$, consistent with the numeral results.

\textcolor{black}{In order to better utilize limited system size and display scrambling dynamics across the chain, we also place the initial operator at the boundary position (site 1). In Fig~\ref{Holevo informationIX_site1}, $\chi_n(\hat{I},\hat{X}_1(t))$ only has half of the light cone, which maximizes the operator spreading distance. This makes the reflections at the system boundaries more clearly visible.}

As a comparison, we also show the spatial-temporal patterns of Holevo information $\chi_n(\hat{X}_4(t),\hat{Y}_4(t))$, as shown in Fig.~\ref{Holevo informationXY}. The patterns are very similar to $\chi_n(\hat{I},\hat{X}_4(t))$, for both $h_z=0$ and $h_z=0.3$. However, a major difference can be seen in the chaotic case, where Holevo information will approach zero. Thus, the initially distinct local operators become indistinguishable eventually with time evolution, making it hard to access the information locally.

The above results are obtained under noiseless quantum simulations. On real quantum hardware, the effect of quantum noises cannot be ignored. In simulation, we use a noise model which adds depolarization to each gate. The noise rate is set as $p=10^{-3}$. We adopt an error-mitigation technique by extrapolating to the zero-noise limit with Richardson's deferred method~\cite{temme_error_2017}. This can be achieved by performing quantum circuits at several different noise rates with a scaling $p_{j}=c_{j}p$, where $p$ is the lowest noise rate feasible on the quantum hardware and $c_{j}$ is the scaling factor. Then a linear combination of results with a weighting $\gamma_{j}$ for the noise rate $p_{j}$ is adopted to make extrapolation to the zero-noise limit. The parameters $c_{j}$ and $\gamma_{j}$ should satisfy two relations,
\begin{equation}
\begin{split}
&\sum_{j=0}^{n_{c}}\gamma_{j}=1,\\
\sum_{j=0}^{n_{c}}&(\gamma_{j}c_{j}^{k})=0~\text{for} ~k=1,2,...,n_c,
\end{split}
\end{equation}
and the error can be reduced to the $O(p^{n_{c}+1})$.  We chose parameters $c_{j}$ and $\gamma_{j}$ for different $n_{c}$ : $c_{0} = 1$, $c_{1} = 2$, $\gamma_{0} = 2$, $\gamma_{1} =-1$ for $n_c = 1$; $c_{0} = 1$, $c_{1} = 2$, $c_{2} = 3$, $\gamma_{0} = 3$, $\gamma_{1} =-3$, $\gamma_{2} = 1$ for $ n_{c} = 2$; $c_{0} = 1$, $c_{1} = 2$, $c_{2}=3$, $c_{3}=4$, $\gamma_{0}=4$, $\gamma_{1}=-6$, $\gamma_{2}=4$, $\gamma_{3}=-1$ for $n_c=3$.

As seen in Fig.~\ref{error}(a) and Fig.~\ref{error}(c), under the influence of noise of a rate $p=10^{-3}$, reasonable in the present quantum hardware, the $\chi_{4}(\hat{X}_4(t),\hat{Y}_4(t))$ of the originally integrable system will appear to decay rapidly, in contrast to the oscillation behavior in the absence of quantum noise. Thus, it is hard to tell an integrable system apart from a chaotic system by Holevo information \textcolor{black}{for the mixed-field Ising model.} By error mitigation to an order of $n_c=3$, one can see that a clear oscillation behavior of Holevo information is restored. Although for a long time, the Holevo information still decays to zero, it is enough to show the distinct behavior between the integrable and the chaotic systems \textcolor{black}{for the mixed-field Ising model.} The behaviors of $\chi_{4}(\hat{I}_4(t), \hat{X}_4(t))$ for the integrable and the chaotic systems are shown in Fig.~\ref{error}(b) and Fig.~\ref{error}(d), respectively. With quantum noise, the Holevo information will saturate to $0.54$ in the long time limit, for both integrable and chaotic systems. With error mitigation, the oscillation behavior for the integrable system will be restored. In this regard, \textcolor{black}{for the mixed-field Ising model,} error mitigation can enable us to reveal the difference between integrable and chaotic systems by measuring the Holevo information.

\section{conclusion}
\label{s5}
In conclusion, we have proposed Holevo information of operators as a measure of quantum operator scrambling by showing the distinct behaviors of the evolution of Holevo information with time for the integrable and the chaotic systems \textcolor{black}{for the mixed-field Ising model.} Moreover, we have developed a quantum algorithm for the effective measurement of Holevo information of operators by using the randomized channel-state duality, where a number of $N+1$ qubits is required. Using the mixed-field Ising model as a demonstration, we have shown with numeral simulations the distinct behavior of spatial-temporal patterns of Holevo information of operators for the integrable and the chaotic systems. While the distinction disappears in the presence of quantum noises, we have found that error mitigation can restore the crucial oscillation feature of Holevo information for the integrable system, making the difference from the chaotic system still observable. One work thus has opened a new perspective as well as a feasible protocol for revealing quantum operator scrambling with Holevo information.

\begin{acknowledgments}
 This work was supported by the National Natural Science Foundation of China (Grant No.12375013) and the Guangdong Basic and Applied Basic Research Fund (Grant No.2023A1515011460).
\end{acknowledgments}
\normalem


\bibliographystyle{apsrev4-1}
\bibliography{Ref3}

\end{document}